\documentstyle[sprocl]{article}

\begin{document} 
 
\title{GREEN'S FUNCTIONS FOR NEUTRINO MIXING} 
 
\author{M. BLASONE${}^{a,*}$, G. VITIELLO${}^{a}$} 
 
\address{${}^{a}$Dipartimento di Fisica and INFN, Universit\`a 
di Salerno,\\ 
84100 Salerno, Italy \\ ${}^{*}$ 
Blackett Laboratory, Imperial College,  
Prince Consort Road,\\London SW7 2BZ, U.K.\\ 
m.blasone@ic.ac.uk\\ 
vitiello@pcvico.csied.unisa.it}

\author{P, A. HENNING} 
 
\address{Institute f\"ur Kernphysik, TH Darmstadt,  
Schlossgartenstrasse 9,\\ 
D-64289 Darmstadt, Germany\\P.Henning@gsi.de} 
 
\maketitle
\abstracts{ The Green's function formalism for  
neutrino mixing is presented and the exact oscillation  
formula is obtained. The usual Pontecorvo formula is  
recovered in the relativistic limit.} 
 
\section{Introduction} 
  
\noindent We report on the Green's 
function formalism for neutrino field mixing recently presented 
in \cite{prl} (see also \cite{BV95,BHV96}). 
The result is an   
oscillation formula which differs from the usual one \cite{BP78} in the  
non-relativistic region. 
We get, together   
with the "squeezing" factor of the amplitude found in ref. 
\cite{BV95},   
also an additional term with a different oscillatory frequency.  
This last feature is   
particularly important since it shows   
that resonance is possible also in vacuum for particular values   
of the masses or of the momentum, thus leading to a suppression or to   
an enhancement of the conversion probability.  
 
We consider two Dirac neutrino fields 
$\nu_e$ and $\nu_\mu$ (space-time dependence   
suppressed). 
The ``flavor mixing'' transformations are 
\begin{eqnarray} \nonumber   
\nu_{e}(x)   &=&\nu_{1}(x)\,\cos\!\theta +    
                        \nu_{2}(x)\,\sin\!\theta\\   
\label{rot1}   
\nu_{\mu}(x) &=&-\nu_{1}(x)\,\sin\!\theta    
                      + \nu_{2}(x)\,\cos\!\theta   
\;,\end{eqnarray}   
where $\theta$ is the mixing angle.  
 $\nu_1$ and $\nu_2$ 
are explicitly given by   
\begin{equation}\label{2.2}\nu_{i}(x) =    
\frac{1}{\sqrt{V}} \sum_{{\bf k},r} 
\left[u^{r}_{{\bf k},i}e^{-i\omega_{k,i} t} \alpha^{r}_{{\bf k},i}\:+    
v^{r}_{-{\bf k},i}e^{i\omega_{k,i} t}\beta^{r\dag }_{-{\bf k},i}  
\right] e^{i {\bf k}\cdot{\bf x}} , \; ~  i=1,2 \;.    
\end{equation}   
We use $t\equiv x_0$, when no misunderstanding  
arises.   
The vacuum for the $\alpha_i$ and $\beta_i$ operators is denoted by   
$|0\rangle_{1,2}$:  
$\; \; \alpha^{r}_{{\bf k},i}|0\rangle_{12}= \beta^{r }_{{\bf    
k},i}|0\rangle_{12}=0$.   
The anticommutation relations,  
the completeness and orthonormality relations are the usual ones. 
 In order to circumvent the difficulty of the   
construction of a Fock space for the mixed fields \cite{Fuj,GKL92},   
it is useful to expand the flavor fields    
$\nu_e$ and $\nu_\mu$ in the same basis as $\nu_1$ and $\nu_2$; e.g.  
$\nu_e$ is given by 
$$
\nu_{e}^{\alpha}(x)     
= G^{-1}(\theta,t)\, \nu_{1}^{\alpha}(x)\, G(\theta,t)   
$$
\begin{equation}
= \frac{1}{\sqrt{V}} \sum_{{\bf k},r}   
  \left[ u^{r}_{{\bf k},1}e^{-i\omega_{k,1}t} \alpha^{r}_{{\bf k},e}(t)
+    
v^{r}_{-{\bf k},1}e^{i\omega_{k,1}t}\beta^{r\dag}_{-{\bf k},e}(t)   
\right]  e^{i {\bf k}\cdot{\bf x}} , 
\end{equation}\label{exnum1}
\noindent where $   
G(\theta,t) =    
\exp\left[\theta \int d^{3}{\bf x}  \left(\nu_{1}^{\dag}(x)\nu_{2}(x) -    
\nu_{2}^{\dag}(x) \nu_{1}(x)\right)\right]\;$,    
is the generator of the mixing transformations (\ref{rot1}) \cite{BV95}.
The flavor annihilation operators can be now explicitly given. 
For example, the electron    
neutrino annihilator is   
$$  
\alpha^{r}_{{\bf k},e}(t) \! = \! \cos\!\theta\,\alpha^{r}_{{\bf k},1} 
\! + \!
$$
\begin{equation}\label{exnue2}
 \sin\!\theta\;\sum_{s}\left(   
u^{r\dag}_{{\bf k},1} u^{s}_{{\bf k},2}   
e^{-i(\omega_{k,2}-\omega_{k,1})t}   
 \alpha^{s}_{{\bf k},2}\,+\,   
u^{r\dag}_{{\bf k},1} v^{s}_{-{\bf k},2}   
e^{i(\omega_{k,1}+\omega_{k,2})t} \beta^{s\dag}_{-{\bf k},2}\right)  \,
. 
\end{equation} 
Notice that it has contributions  
from $\alpha_1$, $\alpha_2$ but {\em also\/}   
from the anti-particle operator $\beta^{\dag}_2$ \cite{BV95} since 
spinor wave   
functions for different masses are not orthogonal. In the more    
traditional   
treatment of mixing, the $\beta^{\dag}_2$ contribution is  
missed since the   
non-orthogonality of the spinor wave functions is not considered.   
   
We can show that when the two point Green's function for the mixed
fields    
$\nu_e$, $\nu_\mu$ are constructed by using the vacuum $|0>_{1,2}$, then
the 
``survival'' probability amplitude, say,  of an electronic    
neutrino    
state in the limit   
$t\rightarrow 0^+$ is computed to be ${\cal P}_{ee}({\bf k},0^+)=   
\cos^2\!\theta + \sin^2\!\theta\, |U_{\bf k}|^2 < 1 $, which is clearly
not acceptable since, of course, it 
should be $\lim_{t\rightarrow {0}^{+}} {\cal   
P}_{ee}(t)=1$. 

Here $|U_{\bf k}|^{2}$ is calculated from the spinor    
basis and its explicit form is given in  
\cite{BV95}.
For different masses and ${\bf k}\neq 0\,$,  
$|U_{\bf k}|$ is always $<1$ \cite{BV95}, and we   
will also use    
$|V_{\bf k}|=\sqrt{1-|U_{\bf k}|^2}$. $|U_{\bf  
k}|^2\rightarrow 1$ in the relativistic limit 
${\bf k}\gg \sqrt{m_1 m_2}$.  
 
The above contradiction shows that 
the {\em choice of the state\/} $|0\rangle_{1,2}$ in   
the computation of the Wightman function   
is {\em not\/}  the correct one.
The problem is in the fact that
the transformation (\ref{rot1}) does not leave invariant 
the vacuum $|0\rangle_{1,2}$. The  
mixing generator induces on it a $SU(2)$ coherent    
state structure, resulting in a new state,   
$|0(\theta,t)\rangle_{e,\mu}\equiv G^{-1}(\theta,t)|0\rangle_{1,2}\,$,   
which is the {\em flavor vacuum\/} for the    
flavor operators $\alpha_{e/\mu}$, $\beta_{e/\mu}$ \cite{BV95}:   
$\alpha_{{\bf k},e/\mu}^{r}(t)|0(t)\rangle_{e,\mu}=\beta_{{\bf    
k},e/\mu}^{r}(t)|0(t)\rangle_{e,\mu}=0$. 
An important feature of the flavor    
vacuum $|0\rangle_{e,\mu}$ (and of the relative Fock space) is its    
non-perturbative nature, resulting in the unitary inequivalence with    
the ``perturbative'' vacuum $|0\rangle_{1,2}$, in the infinite volume    
limit \cite{BV95}. Notice  
that the squared modulus of the   
survival probability amplitude 
reproduces the Pontecorvo oscillation formula in the   
relativistic limit.
   
We show below that the correct  
definition of the Green's functions 
is the one which involves the    
non-perturbative vacuum $|0\rangle_{e,\mu}$.

\section{Green's functions for flavor neutrinos} 
   
In the case of $\nu_e \rightarrow \nu_e$    
propagation,  
the relevant Wightman function  
is (we use $x_0=t, y_0=0$)   
$i G^{>\alpha \beta}_{ee}(t,{\bf x};0,{\bf y}) =    
{}_{e,\mu}\langle0|\nu^{\alpha}_{e}(t,{\bf x}) \;   
\bar{\nu}^{\beta}_{e}(0,{\bf y})   
|0\rangle_{e,\mu}$. It can be 
conveniently expressed in terms of anticommutators at different times   
as   
$$    
i G^{>\alpha\beta}_{ee}({\bf  k},t) = 
$$
\begin{equation}\label{gfu2}
\sum_{r}   
\left[u^{r,\alpha}_{{\bf k},1}\,    
\bar{u}^{r,\beta }_{{\bf k},1}   
\left\{\alpha^r_{{\bf k},e}(t),\alpha^{r\dag}_{{\bf k},e}\right\}  
e^{-i\omega_{k,1}t} +  \, 
v^{r,\alpha}_{-{\bf k},1}\,    
\bar{u}^{r,\beta}_{{\bf k},1}   
\left\{\beta^{r\dag}_{-{\bf k},e}(t),\alpha^{r\dag}_{{\bf k},e} 
\right\}e^{i\omega_{k,1}t} \right]. 
\end{equation}\label{gfu3}   
Here  
$\alpha^{r\dag}_{{\bf k},e}$ stands for 
$\alpha^{r\dag}_{{\bf k},e}(0)$. 
The corresponding transition amplitude is 
\begin{equation} \label{pee3}
{\cal P}^r_{ee} ({\bf  k},t)   
=\cos^{2}\!\theta\,   
 + \sin^2\!\theta\,\left[ |U_{\bf k}|^{2} { e}^{-i   
 (\omega_{k,2}-\omega_{k,1}) t}   
     + |V_{\bf k}|^{2} e^{i(\omega_{k,2}+\omega_{k,1}) t}\right].   
\end{equation}
   
We thus find that the probability amplitude is now   
correctly normalized:  $lim_{t\rightarrow 0^+}   
{\cal P}_{ee}({\bf  k},t)=1$, and one can show that  ${\cal P}_{ee}$,  
${\cal P}_{\mu e}$, ${\cal P}_{\bar{\mu} e}$ go to zero in the same  
limit $t\rightarrow 0^+$ . Moreover,  
\begin{equation}\label{cons}   
\left|{\cal P}^r_{ee}({\bf  k},t)\right|^2 +    
\left|{\cal P}^r_{\bar{e}e}({\bf  k},t)\right|^2 +   
\left|{\cal P}^r_{\mu e}({\bf  k},t)\right|^2 +   
\left|{\cal P}^r_{\bar{\mu}e}({\bf  k},t)\right|^2 =1   
\,,\end{equation}   
as the conservation of the total probability requires.  
Notice that in the    
perturbative case, there were only two non-zero amplitudes, i.e. ${\cal    
P}_{ee}$ and ${\cal P}_{\mu e}$.

At time $t=0$ 
the one electronic neutrino   
state is (momentum and spin indices dropped)    
$|\nu_e \rangle \equiv \alpha_{e}^{\dag}|0\rangle_{e,\mu} $.
In this state a multiparticle component is present,    
disappearing in the relativistic limit $k\gg \sqrt{m_1m_2}\,$, where 
the Pontecorvo state is recovered. Its time evolution   
is given by $ |\nu_e (t)\rangle \equiv e^{-iH t} |\nu_e\rangle $  
and in the flavor basis this state is found to be    
\begin{equation}\label{h3}   
|\nu_e(t) \rangle= \left[\eta_{1}(t)\;\alpha_{e}^{\dag} \;   
+ \;\eta_{2}(t)\;\alpha_{\mu}^{\dag} \;+ \;   
 \eta_{3}(t)\; \alpha_{e}^{\dag}\alpha_{\mu}^{\dag}\beta_{e}^{\dag} \;   
+\;\eta_{4}(t)\;\alpha_{e}^{\dag}\alpha_{\mu}^{\dag}\beta_{\mu}^{\dag}   
\right]|0\rangle_{e,\mu} \,.   
\end{equation}   
Here the $\eta(t)$ are coefficient satisfying the normalization    
condition $|\eta_{1}(t)|^{2}\;+\;|\eta_{2}(t)|^{2}\;+\;  
|\eta_{3}(t)|^{2}\;+ \;|\eta_{4}(t)|^{2}\;=\; 1 $.  

Notice that  
$|0\rangle_{e,\mu}$ is not eigenstate of the free Hamiltonian $H$;  
it ``rotates'' under the action of the time   
evolution generator:  
$|0(t)\rangle_{e,\mu} \equiv  e^{-i H_{1,2} t}\;|0\rangle_{e\mu}$.  
In fact one finds $\lim_{V \rightarrow \infty}\;_{e,\mu}
\langle 0\;|\;0(t)\rangle_{e,\mu} = 0$. 
Thus at different times we have unitarily inequivalent flavor   
vacua (in the limit $V\rightarrow \infty$): this is not surprising  
since 
it is   
direct consequence of the fact that  
the flavor states are not mass   
eigenstates and therefore 
the   
Poincar\'e structure of the flavor vacuum is lacking.

Finally, the charge operators are  
$Q_{e/\mu}\equiv \alpha_{e/\mu}^{\dag}\alpha_{e/\mu} -   
\beta^{\dag}_{e/\mu}\beta_{e/\mu}$.  We have  
$\;_{e,\mu}\langle 0(t)|Q_e/\mu|   
0(t)\rangle_{e,\mu}\; =\; 0 \,$ and 
charge conservation is ensured at any time:  
$\langle \nu_e(t)|\left(Q_e\;+\;Q_{\mu}\right)| 
\nu_e(t)\rangle\; = \; 1$.  
The oscillation   
formula for the flavor charges then readily follows 
$$P_{\nu_e\rightarrow\nu_e}({\bf k},t) = $$
\begin{equation}
1 - \sin^{2}( 2 \theta) \left[ |U_{{\bf k}}|^{2} \;    
\sin^{2} \left( \frac{\omega_{k,2} - \omega_{k,1}}{2} t \right)   
+|V_{{\bf k}}|^{2} \;   
\sin^{2} \left( \frac{\omega_{k,2} + \omega_{k,1}}{2} t \right) \right] 
\, ,  
\end{equation}
$$P_{\nu_e\rightarrow\nu_\mu}({\bf k},t) = $$   
\begin{equation}
  \sin^{2}( 2 \theta)\left[ |U_{{\bf k}}|^{2} \;   
\sin^{2} \left( \frac{\omega_{k,2} - \omega_{k,1}}{2} t \right)    
+|V_{{\bf k}}|^{2} \;   
\sin^{2} \left( \frac{\omega_{k,2} + \omega_{k,1}}{2} t \right) \right] 
\, . 
\end{equation}
This result is exact and includes the previous result of momentum 
dependent oscillation amplitude of refs.\cite{BV95}. 
Notice that the additional contribution to the usual oscillation  
formula, does oscillate with a frequency which is the sum of the  
frequencies of the mass components. In the relativistic limit  
$k\gg\sqrt{m_1m_2}$ the   
traditional oscillation formula is recovered.  
  
\section*{Acknowledgements} 
We acknowledge INFN, MURST and   
the ESF Network on ``Topological defects in Cosmology and 
Condensed Matter'' for partially supporting this work. 
  

\end{document}